**Is $MgB_2$ a superconductor?**
**Comment on "Evidence Against Superconductivity in Flux Trapping Experiments on Hydrides Under High Pressure"** [J. E. Hirsch and F. Marsiglio in *J. Supercond. Nov. Mag.* **35**, 3141–3145 (2022)]

E. F. Talantsev[1]*, V. S. Minkov[2], V. Ksenofontov[2], S. L. Bud´ko[3], M. I. Eremets[2]*

[1]*M. N. Mikheev Institute of Metal Physics, S. Kovalevskoy St. 18, Ekaterinburg 620108, Russian Federation*
[2]*Max Planck Institute for Chemistry; Hahn-Meitner-Weg 1, Mainz 55128, Germany*
[3]*Ames National Laboratory, U.S. Department of Energy, and Department of Physics and Astronomy, Iowa State University; Ames, IA 50011, United States*

*Hirsch and Marsiglio, in their recent publication (J. Supercond. Nov. Mag.* ***35****, 3141–3145, 2022), assert that experimental data on the trapping of magnetic flux by hydrogen-rich compounds clearly demonstrate the absence of superconductivity in hydrides at high pressures. We argue that this assertion is incorrect, as it relies on the wrong model coupled with selective manipulations (hide/delete) of calculated datasets and ignores the reference measurements after the release of pressure. A critical examination of the authors´ claim of having performed fitting of experimental data to the model reveals that, in fact, the authors conducted simulations where all free parameters were fixed. Importantly, an application of the Hirsch-Marsiglio model to* $MgB_2$ *leads to the conclusion that it is not a superconductor.*

Hirsch and Marsiglio proposed a model (Equations 5-7 in Ref.[1]) to describe the field-dependent trapped magnetic moment, $m(B_{appl})$, in superconductors under the zero-field cooled (ZFC) protocol. They claimed to have utilized these equations to fit ZFC $m(B_{appl})$ experimental data, which were measured on the highly-compressed $H_3S$ superconductor[2,3]. Based on the fact that their model failed to accurately describe the experimental $m(B_{appl})$ data, the authors[1] concluded that the measured data do not align with the expected behaviour of magnetic flux trapping in superconductors. Consequently, the authors[1] asserted that $H_3S$ is not a superconductor.

It is crucial to note that the authors[1] did not verify their model using any ZFC or FC data measured in well-studied superconductors. We emphasize that employing an unverified model to describe a physical property or phenomenon or to refute its existence is not a common or standard approach in scientific research.

To address the authors' assertion[1], Bud´ko *et al.* conducted measurements of $m(B_{appl},T)$ data under ZFC conditions in the well-studied superconductor $MgB_2$,[4] following the same protocol used in experiments on $H_3S$ and $LaH_{10}$.[2,3,5] In Figure 1, we present the ZFC $m(B_{appl})$ dataset measured in $MgB_2$ single crystal[4] and the data fit using the Hirsch-Marsiglio model. It is evident that the fit is remarkably poor and cannot adequately describe the experimental data. Following the logic of Hirsch and Marsiglio[1] manifested in Ref.[1], this would mean that $MgB_2$ is not a superconductor.

Significantly, the Hirsch-Marsiglio model does not describe the Meissner regime:

$$m(B_{appl},T) \neq 0, for\ B_{appl} < \mu_0 H_p \quad (1)$$

Instead, the model predicts a significant positive magnetic moment for ZFC $m(B_{appl}=0,T)$ (see Figure 1):

$$m(B_{appl}=0, T=1.8\ K) \cong \frac{1}{7} \times m(B_{appl}=6\ Tesla, T=1.8\ K) \quad (2)$$

This feature of the proposed model was not discussed by the authors[1] and obviously contradicts the physics for the trapped magnetic flux in superconductors.

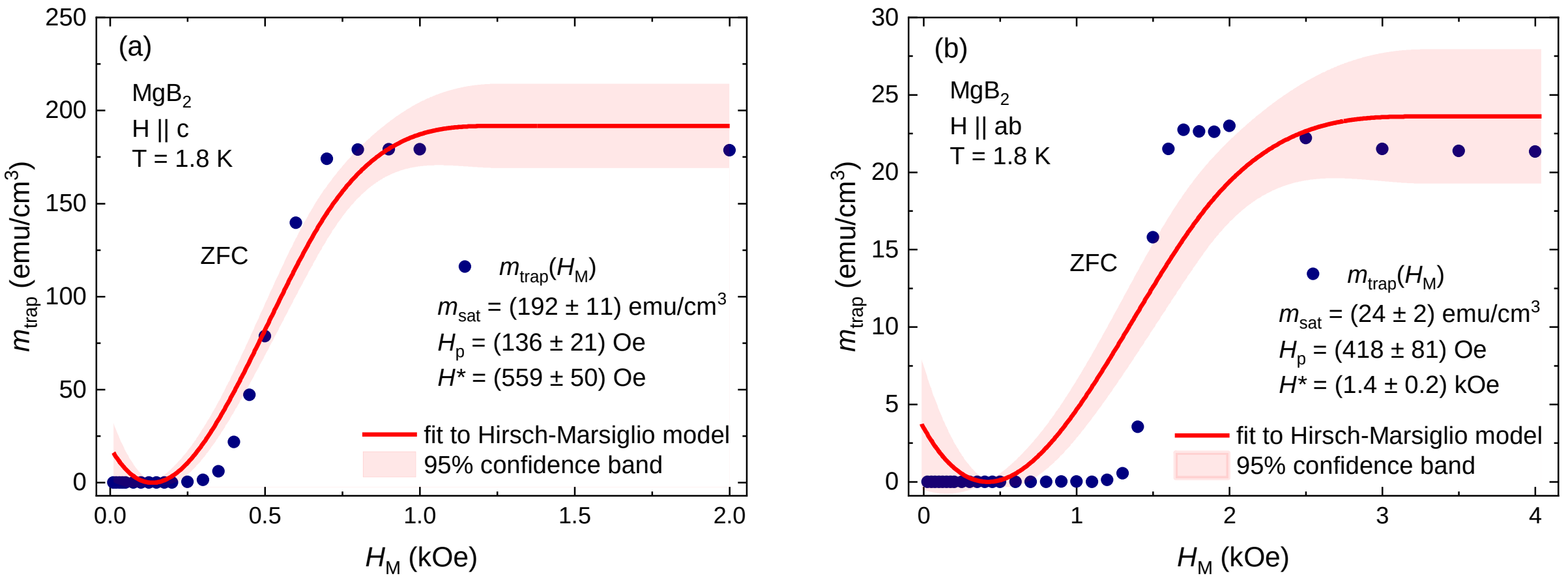

**Figure 1.** Experimental ZFC $m(B_{appl}, T = 1.8\ K)$ datasets measured for $MgB_2$ single crystal[4] and fits of the data by the Hirsch-Marsiglio model proposed in Ref.[1] (a) $m(B_{appl}||c, T = 1.8\ K)$ data; (b) $m(B_{appl}||ab, T = 1.8\ K)$.

Although it is already evident that the Hirsch-Marsiglio model fails to describe the magnetic field-dependent trapped magnetic flux in the well-known superconductor, we nevertheless attempted to reproduce the fitting of our experimental data on $H_3S$ performed by the authors[1]. The FORTRAN 77 fitting code, as provided by Hirsch and Marsiglio, is included in Supplementary Materials. We also converted the code to the more suitable MatLab language used in the present work (see Supplementary Materials). We also used the Origin software to fit/calculate data and prepare figures in this paper.

Surprisingly we discovered that the provided code[1] is a ***simulation*** code, where all four fitting parameters ($\mu_0 H_p$, $\mu_0 H^*$, $m_{sat}$, and the maximal applied field, $\mu_0 H_{max}$ - a parameter not mentioned in Ref.[1]) are fixed and need to be entered manually. The authors mentioned the $\mu_0 H_M^{sat}$ parameter in several places of the text in Ref.[1] but did not use it in the code. In addition, the authors[1] did not provide details on why they chose these specific fixed values for all parameters. Astonishingly, despite claiming that our initial model is incorrect, the authors[1] used the fixed value of $\mu_0 H_p$ determined in our original paper[3].

It is essential to emphasize that the authors[1] ***did not fit*** the experimental data. Instead, they ***simulated*** the $m(B_{appl})$ curves by implementing the following step-by-step procedure:

1. manually choose/set input parameters: $\mu_0 H^*$, $\mu_0 H_p$, $\mu_0 H_{max}$ (entered in the file *input.csv*);
2. manually choose/set $m_{sat}$ (entered in the *cylinder.m* code). In the code provided by the authors[1], $m_{sat} \equiv 1$;
3. generate the $m(B_{appl})$ dataset (using the *usage.m* code);
4. make a *visual assessment of the simulated* $m(B_{appl})$ curve *vs. experimental* $m(B_{appl})$ dataset;
5. repeat *visual assessments of the simulated* $m(B_{appl})$ *curve* (for $B_{appl} \geq \mu_0 H_p$) until it meets the undescribed criteria for visual satisfaction by the authors[1];
6. *manually hide/delete all simulated points* within the Meissner state, i.e. for $B_{appl} \leq \mu_0 H_p$;
7. plot the partially deleted *simulated* $m(B_{appl})$ *curve* together with experimental data and claim that this curve represents the *fitted curve*.

To visualize the authors´ manipulations, we present the full ***simulated curve*** reproduced using the original Hirsch-Marsiglio code and the reported fixed values of $m_{sat} = 15.9\ nAm^2$, $\mu_0 H_p = 42\ mT$, and $\mu_0 H^* = 0.2\ T$, alongside the original Figure 5 from Ref.[1] in Figure 2. It is evident that for all $B_{appl} < \mu_0 H_p$, the simulated curve does not show the Meissner state, namely $m(B_{appl}, T) = 0$ for all $B_{appl} < \mu_0 H_p$.

In Figure 2c, we also present the standard ***fit*** of the experimental $m(B_{appl})$ data to the Hirsch-Marsiglio model. For the fitting, we set $m_{sat}$, $\mu_0 H_p$, and $\mu_0 H^*$ as free-fitting parameters and use the full $m(B_{appl})$ dataset.

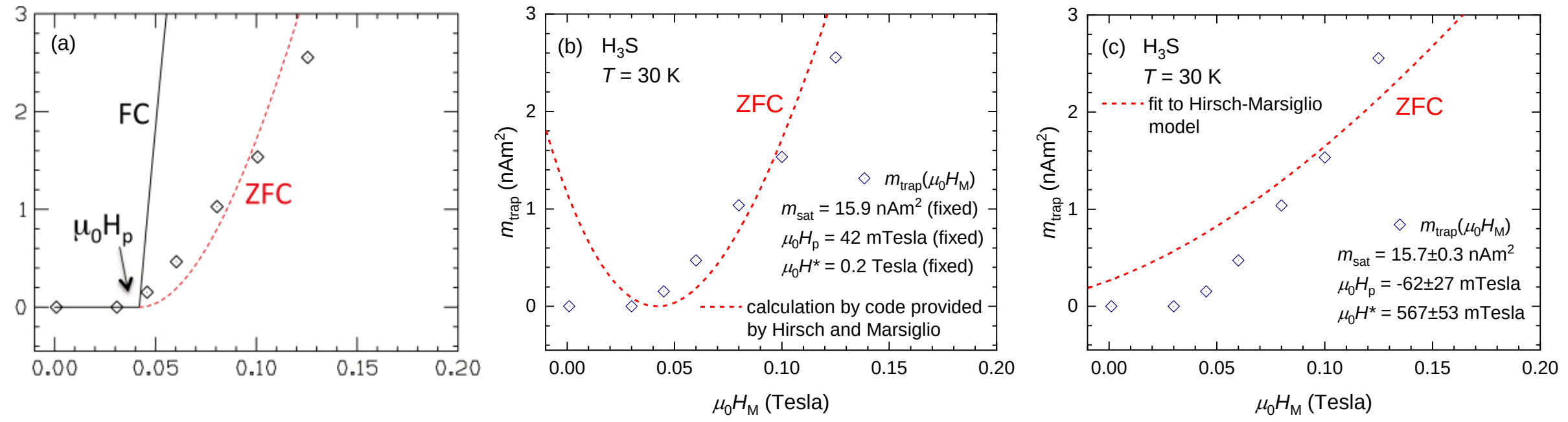

**Figure 2.** Experimental ZFC $m(B_{appl}, T = 30\ K)$ dataset measured in highly compressed $H_3S$[3,4], together with reproduced simulations and standard data fitting. (a) The original Figure 5 from Ref.[1]; (b) the reproduced full simulation curve utilizing the original simulation code, the reported fixed $\mu_0 H_p$, $\mu_0 H^*$, $m_{sat}$, and $\mu_0 H_{max}$ parameters and the aforementioned authors´ manipulations demonstrating the unphysical anomaly within the Meissner state; (c) the standard data fitting to the Hirsch-Marsiglio model where all these parameters are set to be refined. Open rhombuses represent experimental data, black and red curves represent simulation and fitting curves.

In addition to the unphysical issues evident in the standard fitting of experimental data of $MgB_2$ to the Hirsch-Marsiglio model in Figure 1, we found that the resulting refined value for the penetration field is negative, $\mu_0 H_p = -(61 \pm 27)\ mT$. This is an additional indication that the Hirsch-Marsiglio model is incorrect and raises concerns about the presentation of the paper[1] in a way that misleads readers, as the authors[1] did not perform ***data fit*** as claimed in the paper[1].

From all the above, we conclude that the authors[1] in their Figures 3-5[1] ***hid/deleted*** (without reporting this) parts of their simulated datasets that disagree with the Meissner state. This potentially allowed them to conceal the issue that their model and computer code do not adequately describe the Meissner state.

The authors[1] claimed that the measured data either stem from *«experimental artefacts or originate in magnetic properties of the sample or its environment unrelated to superconductivity»*. We argue that the authors overlooked reference measurements conducted on evidently non-superconducting samples, which demonstrate the absence of such artefacts stemming from the sample environment (diamond anvil cells, rhenium gaskets, etc.). The reference measurements revealed that the temperature-dependent magnetic moment of the same samples within the same diamond anvil cells, after the release of pressure, does not exhibit the trapping of magnetic flux. We also note that the authors used the wrong sign for the critical current density, $j_c$, in their Figure 2 for both FC and ZFC panels in Ref.[1] Details can be found elsewhere[6].

It is crucial to emphasize that for any new model, the standard scientific approach includes testing the model in a canonical way: if the deduced parameters of the new model demonstrate good agreement with either the previous model or independent experimental measurements for multiple samples of well-studied compounds, the model can be used for newly discovered compounds. However, even then, it cannot be used as the single exclusive criterion.

Contrary to this standard, the approach implemented by Hirsch and Marsiglio in Ref.[1] does not meet these criteria. As demonstrated in Figure 1 and 2, the Hirsch-Marsiglio model fails to deduce the penetration field and other parameters through standard data fitting. The simulations conducted by the authors[1] involve unverified models, fixing fitting parameters without proper argumentation, and unjustifiably deleting parts of the simulation dataset. Therefore, these simulations cannot be considered as robust "*Evidence Against Superconductivity in Flux Trapping Experiments on Hydrides Under High Pressure*". As a counterexample, we demonstrated that the approach undertaken by the authors[1], which led them to claim the absence of superconductivity in $H_3S$, also implies that $MgB_2$ is not a superconductor.

## Supplementary Materials

**Is $MgB_2$ a superconductor?**
**Comment on "Evidence Against Superconductivity in Flux Trapping Experiments on Hydrides Under High Pressure"** [J. E. Hirsch and F. Marsiglio in *J. Supercond. Nov. Mag.* **35**, 3141–3145 (2022)]

E. Talantsev[1]*, V. S. Minkov[2], V. Ksenofontov[2], S. L. Bud´ko[3], M. I. Eremets[2]*

**1. FORTRAN 77 code for model proposed by J. E. Hirsch and F. Marsiglio (*J. Supercond. Nov. Mag. 35, 3141–3145 (2022)*).**

---

```
c
c       calculates moment for cylinder for FC (unit 11) and ZFC (unit 12)
c       assumes constant jc
c

        implicit real*8 (a-h,o-z)
c
        write (*,*) 'hstar, hp? (1.67,0.042)'
        read (*,*) hstar,hp
        write (*,*) 'hmax? (6)'
        read (*,*) hmax
c
        nh=100
        dh=hmax/nh
        amsat=1.
c
        do ih=0,nh
        hm=dh*ih
        r=1.-hm/hstar
c
        if (r.ge.0)am=amsat*(1.-r**3)
c
        write (11,*) hm,am
        r1=1.-(hm-hp)/hstar/2.
        r2=1.-(hm-hp)/hstar
        if (r2.lt.0.) r2=0.
        if (r1.ge.0.)
     1  am2=amsat*(1.-2.*r1**3+r2**3)
        write (12,*) hm,am2
c
1       continue
        end do
c
        write (11,*) 'join'
        write (12,*) 'join'
        end
```

-----------------------------------------------------------------------------------------------------------

**2. Interpretation of the FORTRAN 77 code used by J. E. Hirsch and F. Marsiglio [*J. Supercond. Nov. Mag.* 35, 3141–3145 (2022)] into MatLab code**

-----------------------------------------------------------------------------------------------------------

**File 1: File name**: usage.m
**Code script:**

```
data = readmatrix('input.csv');
[FC, ZFC] = cylinder(data(1), data(2), data(3));
writematrix(FC, 'FC.csv');
writematrix(ZFC, 'ZFC.csv');
```

-----------------------------------------------------------------------------------------------------------

**File 2: File name**: cylinder.m
**Code script:**

```
function [FC, ZFC] = cylinder(hstar, hp, hmax)
   nh=500;
   dh=hmax/nh;
   amsat=15.9;
   for ih = 0:nh

      hm=dh*ih;
      r=1-hm/hstar;

      if (r>=0)
         am=amsat*(1-r^3);
         FC(ih+1,1) = hm;
         FC(ih+1,2) = am;
      end

      r1=1-(hm-hp)/hstar/2;
      r2=1-(hm-hp)/hstar;

      if (r2<=0)
         r2=0;
      end

      if (r1>=0)
         am2=amsat*(1-2*r1^3+r2^3);
         ZFC(ih+1,1) = hm;
         ZFC(ih+1,2) = am2;
      end

   end

end
```

-----------------------------------------------------------------------------------------------------------

**File 3: File name**: input.csv
File content:

| | | |
|---|---|---|
| 1.67 | 0.042 | 6 |

where, the first column is $\mu_0H^*$ (in Tesla), the second column is $\mu_0H_p$ (in Tesla), and the third column is $\mu_0H_{max}$ (in Tesla).

-----------------------------------------------------------------------------------------------------------

These three files (usage.m, cylinder.m, input.csv) should be in the same folder.